\documentclass[preprint2]{aastex}

\newcommand{\beq}{\begin{equation}}
\newcommand{\eeq}{\end{equation}}
\newcommand{\bea}{\begin{eqnarray}}
\newcommand{\ena}{\end{eqnarray}}
\begin{document}

 \title{On the interpretation of the burst structure of GRBs}

\author{Remo Ruffini, Carlo Luciano 
Bianco, Federico Fraschetti, She-Sheng 
Xue}
\affil{ICRA - International Center for Relativistic Astrophysic\\ Physics 
Department, University of Rome ``La Sapienza''}
\affil{Piazzale Aldo Moro 5, I-00185 Rome, Italy.}
\and
\author{Pascal Chardonnet}
\affil{Universit\'e de Savoie - LAPTH LAPP - BP110 - 74941 Annecy-le-Vieux 
Cedex, France}
\altaffiltext{1}{ruffini@icra.it}
%\altaffiltext{5}{chardon@lapp.in2p3.fr}

\begin{abstract}

Given the very accurate data from the BATSE experiment and RXTE and Chandra satellites, 
we use the GRB 991216 as a prototypical case to test the EMBH theory linking the origin of the energy of GRBs to the electromagnetic energy of black holes.  The fit of the afterglow fixes the only two free parameters of the model and leads to a new paradigm for the interpretation of the burst structure, the IBS paradigm. It leads as well to a reconsideration of the relative roles of the afterglow and burst in GRBs by defining two new phases in this complex phenomenon: a) the injector phase, giving rise to the proper-GRB (P-GRB), and b) the beam-target phase, giving rise to the extended afterglow peak emission (E-APE) and to the afterglow. Such differentiation leads to a natural possible explanation of the bimodal distribution of GRBs observed by BATSE. The agreement with the observational data in regions extending from the horizon of the EMBH all the way out to the distant observer confirms the uniqueness of the model.
\end{abstract}


The most decisive tool in the identification of the energetics of GRBs has been the discovery by Beppo SAX of the afterglow phenomenon. We show in this letter how the afterglow data can be fit using the theory which relates the GRB energy to the extraction process of the electromagnetic energy of a black hole endowed with electromagnetic structure (an EMBH). This energy extraction process occurs via vacuum polarization pair creation and approaches almost perfect reversibility in the sense of black hole physics \citep{cr71,dr75,prx98}.

\begin{figure}
\epsscale{1.0}
\plotone{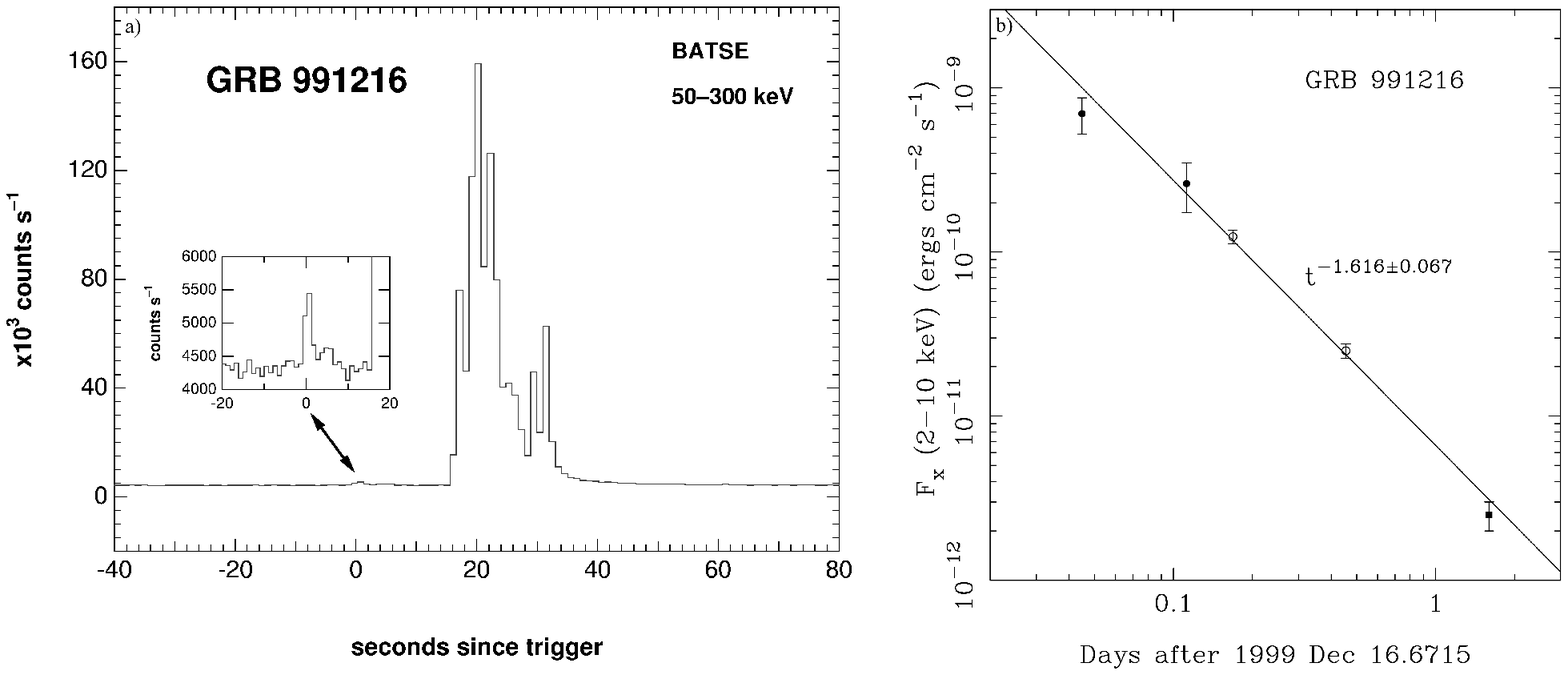}
\caption{a) The data on the GRB~991216 obtained by BATSE (reproduced from BATSE 1999) and b) the corresponding data for the afterglow from both RXTE and Chandra (the last point after $10^5$ s) are given as a function of the detector arrival time \citep[reproduced from][]{ha00}.}
\label{data}
\end{figure}

In addition to yielding excellent agreement between the theory and the data, a new paradigm will be introduced here for the interpretation of the burst structure which we call the IBS paradigm. Because of the unique accuracy of its data, we use the GRB~991216 as a prototype for a description which may then be generalized to other GRBs. The relevant data for GRB~991216 are reproduced in Fig.~\ref{data}, namely the data on the burst as recorded by BATSE \citep{brbr99} and the data on the afterglow from the RXTE satellite \citep{cs00} and the Chandra satellite \citep{p00} \citep[see also][]{ha00}. We have modeled the afterglow assuming that the ultra-high energy baryons (the ABM pulse of \citet{lett1}, Letter 1), accelerated in the pair-electromagnetic-baryonic pulse (PEMB pulse) following a black  hole collapse process (see Letter 1), after reaching transparency interact with the interstellar medium (ISM), assumed to have an average density $n_{ism}$ of 1 proton$/cm^{3}$. All internal energy developed in the collision is assumed to be radiated away in a ``fully radiative'' regime \citep{bfrx01}.  

In our model there are only two free parameters characterizing the EMBH: the mass $M$ in solar mass units $ \mu = M/M_\sun $ and the charge to mass ratio $ \xi = Q/(M\sqrt{G})$, where $M$ and $Q$ are the mass-energy and charge of the EMBH and $G$ is Newton's gravitational constant. These two quantities are related to the total energy of the dyadosphere $\left(E_{dya}\right)$ through the EMBH mass-energy formula \citep{cr71,prx98} as follows
\begin{equation}
\displaystyle{
E_{dya} = \frac{Q^2}{2 \; r_{+}} \left(1 \; - \; \frac{r_{+}}{r_{ds}} 
\right)
\left[ 1 \; - \; \left(\frac{r_{+}}{r_{ds}}\right)^2 \right]\ ,
}
\label{edya}
\end{equation}
where  $r_{+} =  1.47 \times 10^5 \mu ( 1 \;+ \; \sqrt{1 - \xi^2} ) $ is 
the horizon radius and  $ r_{ds} = 1.12 \times 10^8 \sqrt{\mu \xi} $ is the
dyadosphere radius. This energy is the source of the burst $\left(E_{burst}\right)$ and afterglow $\left(E_{aft}\right)$
energies 
\begin{equation}
E_{dya} = E_{burst} + E_{aft}\ .
\label{edya2}
\end{equation}

The only remaining free parameters describe the amount and location of the baryonic matter left over in the collapse process of the precursor star of initial radius $\sim 10^{10}$ cm which forms the EMBH, see \citet{rswx00}. The amount of baryonic matter can be parametrized by the dimensionless parameter $B = (M_Bc^2)/E_{dya}$. As discussed in the previous Letter 1, the results are quite insensitive to the actual density of the baryonic component but they are very sensitive to the value of $B$ \citep{rswx00}. 

\begin{figure}
\epsscale{0.7}
\plotone{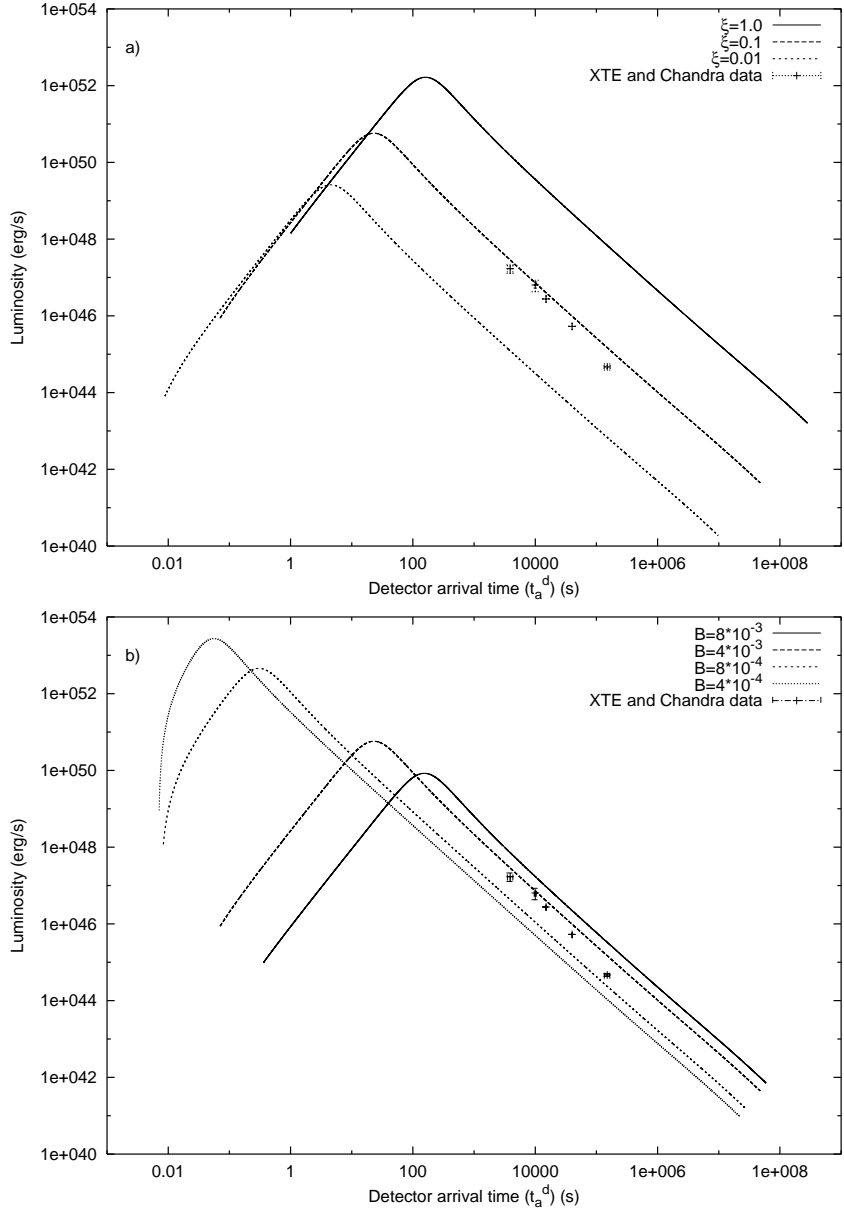}
\caption{a) Afterglow luminosity computed for an EMBH of $22.3M_\sun$ and $B=4\times 10^{-3}$ for three selected values of the electromagnetic parameter $\xi=0.01,0.1,1.0$. b) for the same EMBH mass and $\xi=0.1$, we give the afterglow luminosities corresponding respectively to $B=4\times 10^{-4}, 8\times 10^{-4}, 4\times 10^{-3}, 8\times 10^{-3}$.}
\label{xib}
\end{figure}

In Fig.~\ref{xib} we present some of the results of fitting the data from the RXTE and Chandra satellites corresponding to an EMBH mass of 22.3 $M_\sun$ and for selected values of the parameters $\xi$ and $B$. The main conclusions from our model are the following:

1) The slope of the afterglow, $n=-1.6$, is rather insensitive to the values of the parameters $\mu$, $\xi$ and $B$ and is in perfect agreement with the observational data. The physical reason for this universality of the slope is essentially related to the ultra-relativistic energy of the baryons in the ABM pulse, the assumption of constant average density in the ISM, the ``fully radiative'' conditions leading predominantly to the X-ray emission, as well as all the different relativistic effects presented in the RSTT paradigm. See for details \citet{lett6aa}.

2) The afterglow fit does not depend on all three parameters $\mu, \xi$ and $B$ but only on the combinations $E_{dya}$ and $B$.  Thus there is a 1-parameter family of values of the pair $(\mu,\xi)$  allowed by a given viable value of  $E_{dya}$.

\begin{figure}
\epsscale{1.0}
\plotone{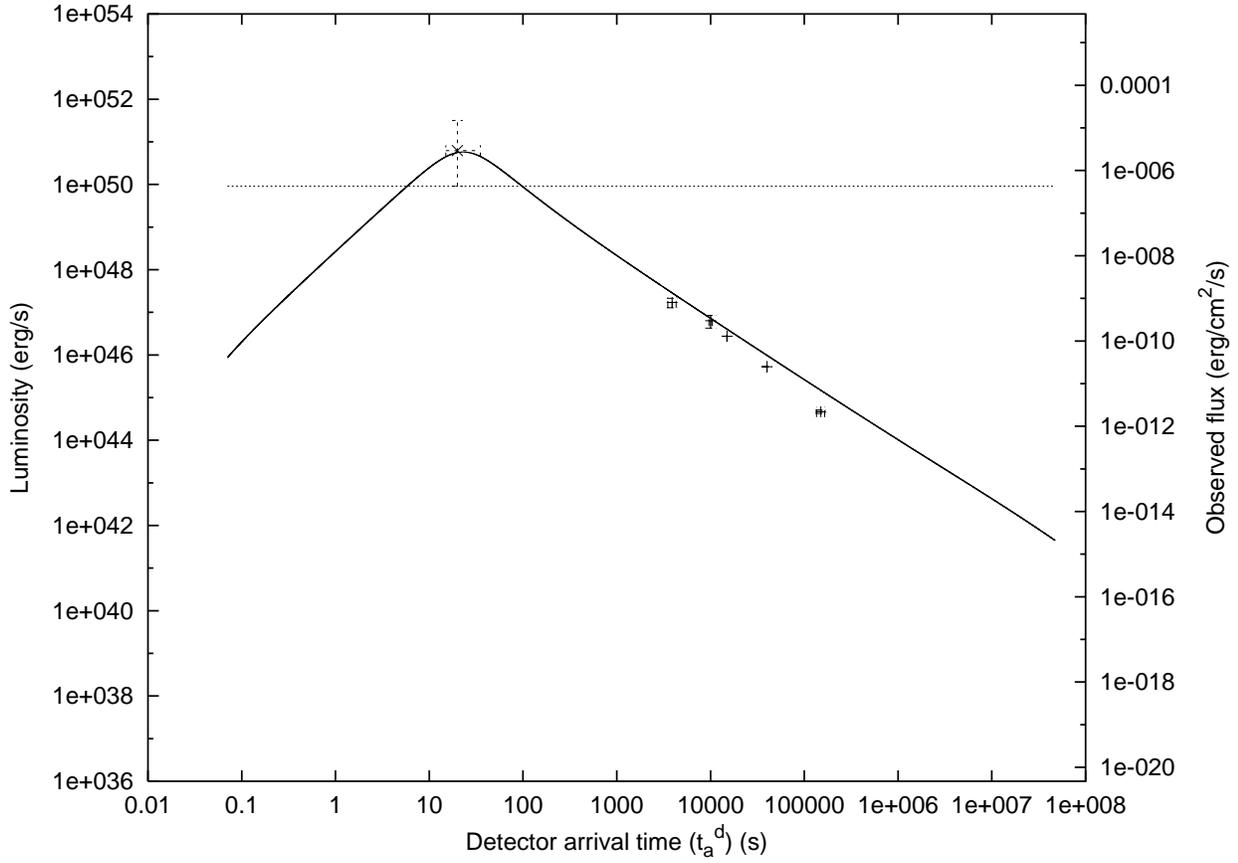}
\caption{Best fit of the afterglow data of Chandra, RXTE as well as of the range of variability of the BATSE data on the major burst, by a unique afterglow curve leading to the parameter values $E_{dya}=9.57\times 10^{52}erg, B=4\times 10^{-3}$. The horizontal dotted line indicates the background noise of this observation. On the left axis the luminosity is given in units of the energy emitted at the source, while the right axis gives the flux as received by the detectors.}
\label{fit}
\end{figure}

3) It is clear, both from studying the profiles and the time dependence of the afterglow, that by suitably modifying the values of B and $\xi$, the average flux of the main burst observed by BATSE can also be fit by the afterglow curve, up to the degeneracy in $(\mu,\xi)$, leading to (see Fig.~\ref{fit}):

\begin{equation}
E_{dya}=9.57\times 10^{52}erg,\; \; B=4\times 10^{-3} \ .
\label{values}
\end{equation}

The peak of the average afterglow emission occurs at $\sim 23.7$ seconds and its intensity and time scale are in excellent agreement with the BATSE observations, an important result \citep[see also][]{lett5}. In addition to the BATSE data, there is also clearly perfect agreement with the decaying part of the afterglow data from the RXTE and Chandra satellites. It is clear that such an extended afterglow peak emission (E-APE) is {\em not} a burst, but it is seen as such by BATSE because of the background noise level in this observation \citep[see also][]{lett5}. Thus the long lasting unsolved problem of explaining the long GRBs \citep[see e.g.][]{wmm96,swm00,p01} is radically resolved.

After we fix the free parameters of the EMBH theory, modulo the mass-charge relationship which fixes $E_{dya}$, all other features of the observations must be explained by the theory.

\begin{figure}
\epsscale{1.0}
\plotone{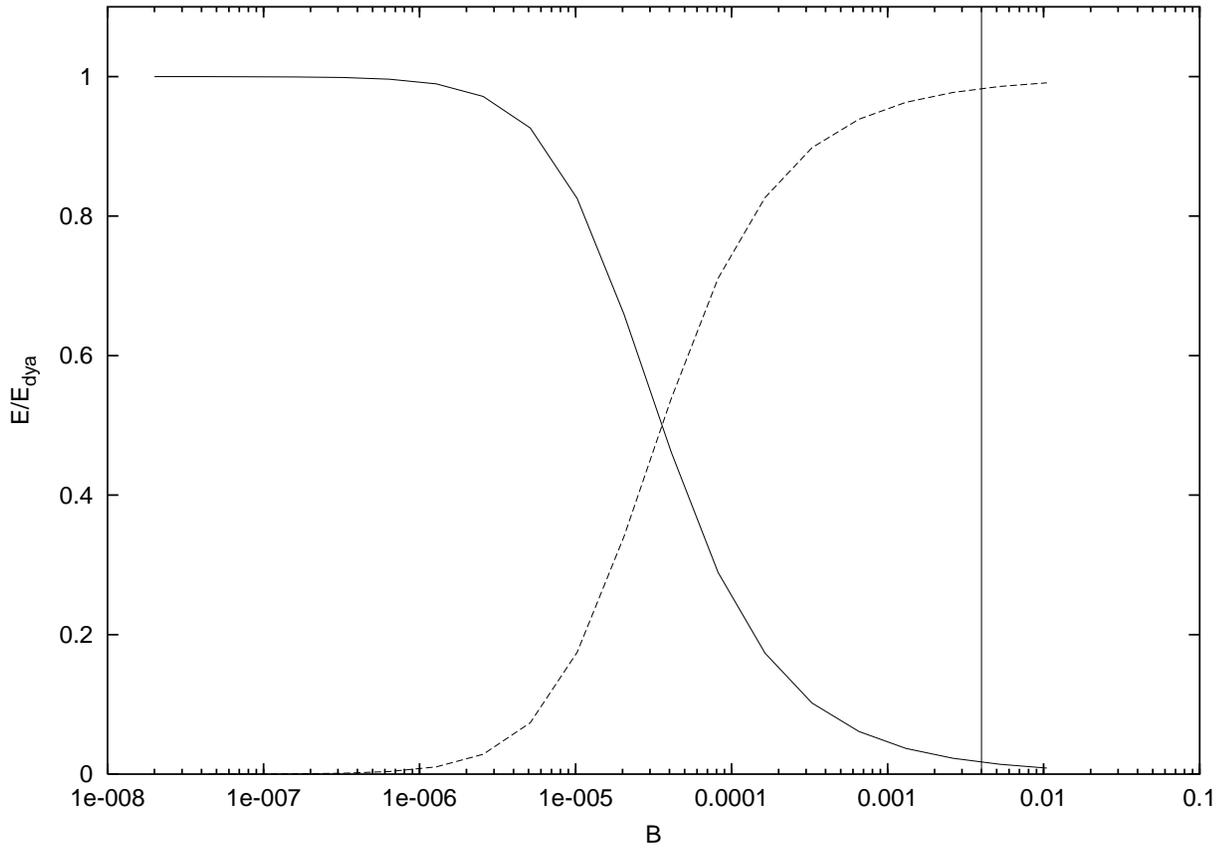}
\caption{Relative intensities of the afterglow (dashed line) and the P-GRB (solid line), as predicted by the EMBH model corresponding to the values of the parameters determined in the previous Fig.~\ref{fit}, as a function of $B$. Details are given in \citep{bfrx01}. The vertical line corresponds to the value $B=4\times 10^{-3}$.}
\label{crossen}
\end{figure}

There is a natural question to be asked: where does one find the burst which is emitted when the condition of transparency against Thomson scattering is reached? We refer to this as the proper-gamma ray burst (P-GRB) in order to distinguish it from the global GRB phenomena (see Letter 1 and \citet{brx00}). \citet{rswx00} showed that, for a fixed value of $ E_{dya}$, a value of $B$ uniquely determines the energy of the P-GRB, which we indicate by $E_{burst}$, and the energy of the afterglow $E_{aft}$ (see Fig.~\ref{crossen}). For the particular values of the parameters given in Eq.~(\ref{values}), we then predict 
\begin{eqnarray}
\frac{E_{burst}}{E_{dya}}=1.45\times 10^{-2};\nonumber \\
E_{burst}=1.39\times 10^{51}erg; \label{fittedvalues} \\
E_{dya} = 9.57\times 10^{52}erg\ . \nonumber
\end{eqnarray}

Is there any evidence of such a signal in the BATSE data?  From the relative time transformation paradigm presented in Letter 1, we can retrace such a P-GRB by reading off the time parameters of point 4 in Fig.~1 from Tab.~1, both in Letter 1. The transparency is reached at $14.23$ sec in comoving time, at a radial coordinate $r=9.692\times 10^{13}$ cm in the laboratory frame and at $1.361\times 10^{-1}$ sec in arrival time at the detector.  All this, namely the energy predicted in Eq.(\ref{fittedvalues}) for the intensity of the burst and its time of arrival, leads to the unequivocal identification of the P-GRB with the apparently inconspicuous initial burst in the BATSE data. We have estimated the ratio of the first peak (the P-GRB) to the E-APE over the background noise level of the BATSE data to be $\sim 10^{-2}$, in very good agreement with the first entry in Eq.~(\ref{fittedvalues}).

In summary, the observational data agree with the predictions of the model on:\\
1) the intensity ratio, $1.45\times 10^{-2}$,  between the P-GRB and the E-APE, which strongly depends on the parameter $B$,\\
2) the absolute intensities for both the P-GRB and the E-APE, respectively $1.39\times 10^{51}$ erg  and  $9.43\times 10^{52}$, which depends on $E_{dya}$,\\
3) the arrival time of the P-GRB and the peak of the E-APE, respectively $1.361 \times 10^{-1}$ sec and $23.7$ sec.

Without the introduction of any new parameter, the model offers additional information both on the detailed structure of the P-GRB and of the E-APE.

\begin{figure}
\epsscale{1.0}
\plotone{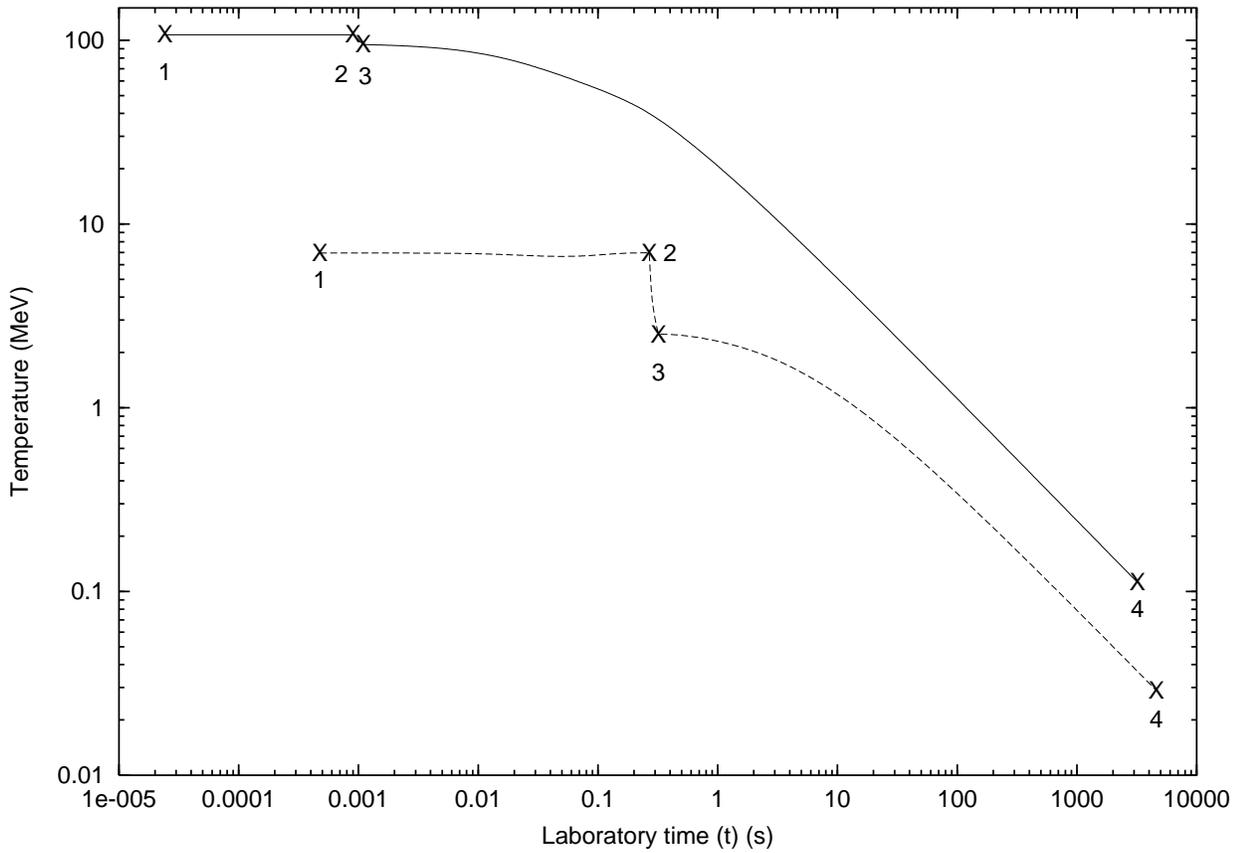}
\caption{The temperature of the pulse in the laboratory frame for the first three eras of Fig. 1 of Letter 1 is given as a function of the laboratory time. The numbers 1,2,3,4 represent the beginning and end of each era. The two curves refer to two extreme approximations adopted in the description of the dyadosphere. Details are given in \citep{rswx00,bfrx01}.}
\label{temp}
\end{figure}

Regarding the P-GRB spectrum, the initial energy of the electron-positron pairs and photons in the dyadosphere for given values of the parameters can be easily computed following the work of \citet{prx98}. We obtain respectively $T=1.95$ MeV and $T=29.4$ MeV in the two approximations we have used \citep{bfrx01}: for a given $E_{dya} $ we have assumed either a constant average energy density over the entire dyadosphere volume, or a more compact configuration with energy density equal to the peak value. It is then possible to follow, in the laboratory frame, the time evolution of the temperature of the electron-positron pairs and photons through the different eras presented in Letter 1, see Fig.~\ref{temp}. The condition of transparency is reached at temperatures in the range of $\sim 15-55$ keV at the detector, in agreement with the BATSE results.

Regarding the E-APE, all the above considerations refer to the smoothed average emission. It is interesting that the detailed structure of the E-APE observed by BATSE can also be reproduced in the model in terms of relativistic effects and deviations from the average value of the ISM density due to inhomogeneities \citep{lett5,lett6aa,rcvx01g}.

We can now proceed to the formulation of the IBS paradigm: in GRBs we can distinguish an {\it injector phase} and a {\it beam-target phase}. The {\it injector phase} includes the process of gravitational collapse of a progenitor star to a black hole endowed with electromagnetic structure (EMBH), the formation of the dyadosphere and the associated phenomena of vacuum polarization as well as the different eras presented in Letter 1: era I corresponds to the PEM pulse, era II to the engulfment of the baryonic matter of the remnant and era III the PEMB pulse. The injector phase terminates at the point where the plasma transparency condition is reached and the P-GRB is emitted. The {\it beam-target phase} addresses the interaction of the ABM pulse, namely the beam generated during the injection phase, with the ISM as the target. It gives rise to the E-APE and the decaying part of the afterglow.

We advance the possibility, to be verified on the basis of the time variabilities and spectral information mentioned above \citep{brx00,rcvx01g}, that the P-GRBs coincide with the class of short events ($<2$ sec) discovered in the bimodal distribution of GRBs in the BATSE catalogue \citep{k93}, while the E-APEs coincide with the class of longer events ($>2$ sec).

It is interesting that, even in this very energetic case of GRB~991216, the general energetic requirement can be easily fulfilled by an EMBH with $M=22.3M_\odot$ and $\xi=0.1$. No beaming is needed, and no evidence of beaming is obtained by the fitting of the theory and the observational data, contrary to views expressed by e.g. \citet{ha00}. See \citet{lett6aa} for details.

As the EMBH model is confirmed by additional sources, the GRBs will be used to scan the regions around the newly formed EMBHs, to infer their physical and astrophysical composition as well as to acquire information on the process of gravitational collapse leading to the EMBH and on the astrophysical structures in the high redshift universe. The first clear intuition of such a possibility has been expressed by \citet{dm99}. We will give a first application of such a ``tomographic'' imaging technique in \citet{lett3}.

We conclude:\\
1) In the range of distances (see Letter 1 Tab. 1) $r\simeq 10^{14}\sim 10^{17}$ cm from the EMBH, information on the ISM, $\bar\rho\simeq 10^{-24} g/cm^3$, and on the surrounding additional astrophysical systems can be inferred from the E-APE and from the afterglow \citep[see][]{lett3,lett5,lett6aa}.\\
2) At a distance $r\simeq 10^{10}$ cm, where $\bar\rho\simeq 1 g/cm^3$ (see Letter 1 Tab. 2), we can evaluate the percentage of mass of the progenitor star left in the remnant by the process of gravitational collapse. We have in fact \citep[see][]{bfrx01}:
\begin{equation}
M_B c^2 = BE_{dya} \simeq \frac{1}{4}B\xi^2c^2M_{BH}
\label{MBsuMBH1}
\end{equation}
which in the case of the GRB~991216 implies that up to 99.9\% of the matter of the progenitor star collapses to the EMBH. This indicates that the gravitational collapse to a black hole differs markedly from the corresponding process occurring in neutron star formation \citep{bfrx01}.\\
3) At $r < r_{ds} \simeq 10^8 cm$ the electrodynamical constraints imply $\bar\rho < 10^{-9} g/cm^3$, to avoid baryonic contamination in the dyadosphere. This condition can be easily satisfied during the gravitational collapse to an EMBH as the horizon is approached. The details of such a process, with all its general relativistic effects, can be followed through the structure of the P-GRB \citep{rcvx01g}.

The IBS paradigm we have introduced is common to a number of models based on a single process of gravitational collapse, leading to GRBs. The uniqueness of the EMBH model resides a) in the energetics \citep{rukyoto}, b) in the time structure of the P-GRB \citep{rcvx01g,rvx01}, c) in the spectral information of the P-GRB \citep{bfrx01}.

The fact that the model is testable from the ISM all the way down to the horizon of the EMBH offers an unprecedented tool for proving its uniqueness, confirming that we are witnessing the formation of EMBHs and the extraction of their electromagnetic energy through the resulting GRBs.

The intrinsic simplicity of the EMBH model of GRBs, shown here to depend only from two parameters, offers an unique opportunity to use GRBs as ``standard candles'' in cosmology.

\acknowledgments

We thank three anonymous referees for their remarks, which have improved the presentation of this letter

\clearpage

\onecolumn

\end{document}